\documentclass[11pt]{article}

\usepackage[margin=1in]{geometry}

\usepackage{mathrsfs}
\usepackage{siunitx}
\usepackage{upgreek}
\usepackage{graphicx}
\usepackage{amsmath}
\usepackage{amssymb}
\usepackage{xfrac}
\usepackage{xcolor}
\usepackage{soul}
\usepackage{setspace}
\usepackage{hyperref}
\hypersetup{
    colorlinks=true,
    linkcolor=blue,
    filecolor=blue,      
    urlcolor=blue,
    citecolor=blue,
   }
\usepackage[normalem]{ulem}

\usepackage{mathptmx}

\usepackage[square, numbers, sort&compress]{natbib}

\usepackage{titlesec}
\titleformat{\section}{\Large\bfseries}{\thesection.}{1em}{}[]
\titleformat{\subsection}{\normalsize\bfseries}{\thesubsection.}{1em}{}[]
\titleformat{\subsubsection}{\normalsize\itshape}{\thesubsubsection.}{1em}{}[]
\titlespacing{\section}{0pt}{5pt}{0pt}[0pt]
\titlespacing{\subsection}{0pt}{3pt}{0pt}[0pt] 
\titlespacing{\subsubsection}{0pt}{0pt}{0pt}[0pt] 

\usepackage[font=footnotesize,labelfont=bf]{caption} 

\usepackage[color=blue!15]{todonotes}
\usepackage{xcolor}

\begin{document}

\begin{center}

\Large
\textbf{Geometry-Driven Mechanical Memory in a Random Fibrous Matrix}
\vspace{11pt}

Mainak Sarkar,$^\mathrm{a,*}$ Christina Laukaitis,$^\mathrm{a,c,e,g}$ Amy Wagoner Johnson$^\mathrm{a,b,c,d,f}$
\vspace{11pt}

\small
$^\mathrm{a}$Carl R. Woese Institute for Genomic Biology, University of Illinois Urbana-Champaign \\ $^\mathrm{b}$ Mechanical Science and Engineering, Grainger College of Engineering, University of Illinois Urbana-Champaign \\ $^\mathrm{c}$Biomedical and Translational Sciences, Carle Illinois College of Medicine, University of Illinois Urbana-Champaign \\  $^\mathrm{d}$Beckman Institute for Advanced Science and Technology, University of Illinois Urbana-Champaign \\ $^\mathrm{e}$Clinical Sciences, Carle Illinois College of Medicine, University of Illinois Urbana-Champaign \\ $^\mathrm{f}$CZ Biohub Chicago, LLC, Chicago, Illinois \\ $^\mathrm{g}$ Carle Health, Urbana, Illinois \\ $^\mathrm{*}$ Corresponding author: Mainak Sarkar, mainak@illinois.edu, 608-213-6684 \\

\end{center}

\section*{Abstract}
Disordered fibrous matrices, formed by the random assembly of fibers, provide the structural framework for many biological systems and biomaterials. Applied deformation modifies the alignment and stress states of constituent fibers, tuning the nonlinear elastic response of these materials. While it is generally presumed that fibers return to their original configurations after deformation is released, except when neighboring fibers coalesce or individual fibers yield, this reversal process remains largely unexplored. The intricate geometry of these matrices leaves an incomplete understanding of whether releasing deformation fully restores the matrix or introduces new microstructural deformation mechanisms.
To address this gap, we investigated the evolution of matrix microstructures during the release of an applied deformation. 
Numerical simulations were performed on quasi-two-dimensional matrices of random fibers under localized tension, with fibers modeled as beams in finite element analysis. After tension release, the matrix exhibited permanent mechanical remodeling, with greater remodeling occurring at higher magnitudes of applied tension, indicative of the matrix preserving its loading history as mechanical memory. This response was surprising; it occurred despite the absence of explicit plasticity mechanisms, such as activation of inter-fiber cohesion or fiber yielding. We attributed the observed remodeling to the gradient in fiber alignment that developed within the matrix microstructure under applied tension, driving the subsequent changes in matrix properties during the release of applied tension. 
Therefore, random fibrous matrices tend to retain mechanical memory due to their intricate geometry.

\section*{Keywords}
fibrous matrix, mechanical memory, microstructural gradient, mechanical remodeling 
\noindent


\newpage
\setcounter{section}{0}
\section*{Introduction}
Fibrous matrices composed of randomly assembled fibers permeate the natural world, appearing in diverse forms such as the cellular cytoskeleton \cite{janmey1998cytoskeleton, bausch2006bottom, alberts2007molecular}, tissue extracellular matrices \cite{muiznieks2013molecular}, and biomaterials like cellulose \cite{klemm2005cellulose}, mycelium \cite{islam2017morphology}, silk proteins \cite{vepari2007silk}, and eggshell membranes \cite{balavz2014eggshell}, among others. When subjected to tension, the fibers bend \cite{licup2015stress, vahabi2016elasticity, picu2018poisson} and align with the direction of maximal principal strain \cite{liang2016heterogeneous, ronceray2016fiber}, accumulating axial stretch \cite{mann2019force, sarkar2022evolution}, while those fibers oriented transverse to the applied deformation undergo axial compression and may buckle \cite{burkel2017mechanical, grimmer2018displacement}. This complex interplay between stretched and buckled fibers, even when the fibers themselves are linear, gives rise to the nonlinear elastic response of the matrix \cite{onck2005alternative, storm2005nonlinear}, manifesting either as strain stiffening \cite{roeder2002tensile, janmey2007negative, brown2009multiscale, vader2009strain, munster2013strain, kim2014structural} or strain softening \cite{sarkar2024unexpected, zakharov2024clots}. This understanding of elasticity has fostered a prevailing assumption in traditional mechanics models of fibrous matrices \cite{grekas2021cells, sopher2018nonlinear, ban2019strong, humphries2017}, with a few recent exceptions \cite{favata2022internal, favata2024emerging}, that fibers revert to their original orientation and zero-stress states once deformation is released. However, given the inherent complexity and intricacy of matrix geometry, the specific microstructural deformation mechanisms likely activated during the release of applied deformation remain elusive.

The intrinsic disorder of random fibrous matrices categorizes them within the broad spectrum of disordered materials \cite{nicolas2018deformation}, which also includes granular systems \cite{guyon1990non,roux2000geometric}. Granular systems preserve mechanical memory of past deformations even after the deformation is released. Specifically, granular materials encode mechanical memory through particle rearrangement and contact network formation during deformation, creating distinct stable patterns that enable the material to respond differently when the deformation is released \cite{keim2019memory,paulsen2024mechanical}. Similarly, fibers in fibrous matrices realign in response to deformation, reflecting behavior observed in granular materials. Given this similarity between fibrous matrices and granular materials, along with other shared characteristics such as tension-compression bimodularity \cite{mann2019force,rosakis2015model} and the development of force chains \cite{sarkar2022evolution,mann2019force,ruiz2022force}, there is a compelling need to rigorously explore whether fibrous matrices also encode mechanical memory through fiber rearrangement. Previous studies on biopolymer fibrous matrices have primarily emphasized memory properties related to traditional plasticity such as the formation of new inter-fiber crosslinks, fiber yielding \cite{kim2017stress,ban2018mechanisms,zhang2019modeling}, or the opening of fiber bundles \cite{burla2020connectivity}. However, they have often overlooked the type of memory derived solely from geometric rearrangements, akin to that observed in granular materials. This gap underscores the need for a more focused investigation into the state of microstructure and its role both during and after the release of applied deformation in fibrous matrices.

In this study, we used mechanics models of a naturally occurring random fibrous matrix, inspired by the collagen structure prevalent in tissues, to investigate mechanical memory formation within matrices. The models here were devoid of traditional plasticity mechanisms like inter-fiber cohesion or fiber yielding. Recognizing that natural fibrous matrices typically experience localized tension from cellular forces, we examined whether these matrices could retain the history of applied cellular forces even after the forces were released. To simulate this process, we embedded an inclusion within the matrix model that, akin to a cell, contracted to exert tension on the matrix. We then allowed the inclusion to recover its original size in order to release the tension. We analyzed the kinematic and stress states of the constituent fibers as the applied tension was released, revealing an unexpected mechanical remodeling of the matrix that is indicative of the inherent ability of the matrix to form and retain mechanical memory of past deformation. This response was linked to a geometric gradient in the matrix microstructure that formed during the application of deformation through the contracting inclusion, where distinct geometric patterns emerged, and fundamentally altered the mechanical response of the matrix during the release of applied deformation as the inclusion recovered. Thus, random fibrous matrices inherently possess the ability to maintain mechanical memory, driven solely by their complex geometry.

\section*{Methods}
\subsection*{Numerical analysis of inclusion-matrix models}
\begin{figure}[t]
\centering
\includegraphics[width=3.3in, keepaspectratio=true]{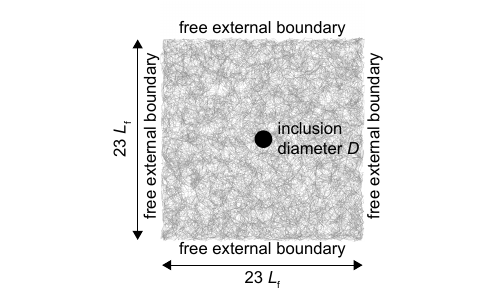}
\caption{Representative matrix containing an embedded contracting inclusion with a diameter ($D$) twice the average fiber length ($L_\mathrm{f}$). The inclusion undergoes radial contraction while the external edges of the matrix remain free.}
\label{fig1}
\end{figure}

We used quasi-two-dimensional discrete fiber models of fibrous matrices with randomly organized fibers, incorporating a contractile circular inclusion embedded within the matrix (Fig. \ref{fig1}). This model replicated the structural intricacies of three-dimensional collagen in two dimensions, while the embedded contractile inclusion emulated typical cellular contraction, mimicking cell-induced mechanical tension on the matrix. This approach follows a modeling strategy that has been established over the past decade (e.g., Ref. \cite{notbohm2015microbuckling}). The matrix model was initially developed elsewhere \cite{lindstrom2010biopolymer, lindstrom2013finite} and was later optimized for numerical efficiency \cite{grimmer2018displacement, proestaki2021effect, sarkar2024unexpected}. These quasi-two-dimensional matrix models have been shown to effectively capture the evolving microstructure and nonlinear mechanics of three-dimensional fibrous matrices \cite{grimmer2018displacement, proestaki2021effect, sarkar2024unexpected}, demonstrating the broad applicability of such models as reported in other studies \cite{licup2015stress, notbohm2015microbuckling, goren2020elastic, hatami2021mechanical, sarkar2022evolution}. 

Mimicking the random structure of collagen matrices, fibers were randomly assembled within a two-dimensional domain to achieve a desired average fiber length ($L_\mathrm{f}$), while ensuring an average nodal connectivity of fibers at $3.4$ \cite{lindstrom2010biopolymer, arzash2019stress}, which is below Maxwell’s isostatic threshold of twice the system dimensionality \cite{maxwell1864calculation}. Similar to collagen and other biological matrices \cite{stein2008algorithm, burkel2018heterogeneity}, this quasi-two-dimensional model allowed fibers to cross each other without connecting, thus balancing computational efficiency with structural fidelity. In finite element simulations, matrix fibers were modeled as linear elastic Timoshenko beams with a shearing stiffness set to half the axial stiffness and a bending-to-axial stiffness ratio of $1 \times 10^{-4}$, typical of collagen \cite{licup2015stress, vahabi2016elasticity, feng2015alignment, van2016uncoupling}. Following prior work \cite{sarkar2022evolution, sarkar2023bioinspired, sarkar2024unexpected}, each fiber was represented by two three-node quadratic beam elements to accurately capture bending and buckling effects. Fibers transmitted force and moment through welded nodes. To avoid any plasticity mechanisms, interactions between fibers were non-cohesive, and it was ensured that fibers under tension did not yield. The external dimensions of the matrix models were $23L_\mathrm{f}$. 

A circular inclusion with diameter $2L_\mathrm{f}$ was embedded using established methodologies \cite{islam2019random,sarkar2024unexpected}; see Supplemental Note 1 for the rationale behind choosing this specific inclusion diameter. Specifically, fibers intersecting the inclusion periphery were trimmed, while those passing through remained intact to preserve structural fidelity akin to a three-dimensional matrix. The inclusion was composed of three-node continuum triangle elements, and two-node linear beam cross-links rigidly connected it to the matrix. Both the inclusion and the cross-links had stiffness values ten orders of magnitude greater than the axial stiffness of the matrix fibers, ensuring that the strain in each inclusion matched the actual strain on the matrix. 

All finite element simulations were conducted in Abaqus (Dassault Systèmes) using two load steps. The first step induced thermal contraction of the inclusion, while in the second step, the inclusion thermally recovered to its original size. The external matrix boundaries remained unconstrained in both steps. The inclusion was modeled as a continuum linear elastic body that maintained both force equilibrium and kinematic compatibility during thermal contraction and recovery, consistent with the methodology described in Ref. \cite{sarkar2024unexpected}. The measure adopted for the radial strain ($\epsilon_\mathrm{r}$) of the inclusion was engineering strain, in accordance with standard practices in previous studies of fibrous matrices \cite{islam2018effect,sarkar2022evolution}. All length dimensions, including references to sizes of selected regions of interest in the matrix, were defined in the Lagrangian coordinate system.

An implicit dynamic quasi-static solver was used with the nonlinear geometry option enabled to account for large local deformations and mechanical instabilities within the matrix, as described in prior studies \cite{proestaki2021effect, sarkar2022quantification, sarkar2023bioinspired, sarkar2022evolution, sarkar2024unexpected}. To ensure accurate convergence of the quasi-static solver, we maintained an underdamped system, following recommendations in Refs. \cite{natario2014web, sarkar2023bioinspired}. All simulations were repeated across eight independent inclusion-matrix models, with each model generated using a different random seed to assemble the fibers in the matrix.

\subsection*{Characterization of matrix mechanics}
\subsubsection*{Measuring buckling-mediated defects}
Matrix mechanics were characterized by measuring the extent of fiber buckling, referred to in this study as defects. During inclusion contraction, fiber buckling was quantified as the excess length of individual fibers ($\xi_\mathrm{f}$) quantified as the difference between the fiber’s contour and its node-to-node distances, as previously described \cite{licup2015stress,vahabi2016elasticity,sharma2016strain,sarkar2022evolution,sarkar2024unexpected}. The spatial evolution of defects with distance $r$ from the inclusion boundary was determined by calculating the areal density of defects ($\psi_{\mathrm{1}}$) at each $r$, achieved by summing the defects of all fibers ($\sum \xi_{\mathrm{f}}$) within thin annular regions of thickness $\delta{r}=0.5L_{\mathrm{f}}$ (Fig. \ref{fig2}a) and dividing by the area of these regions. During inclusion recovery, defects induced in individual fibers were quantified as excess lengths induced ($\Delta\xi_\mathrm{f}$), with the areal density ($\Delta\psi$) at each $r$ calculated by summing these defects across all fibers ($\sum \Delta\xi_\mathrm{f}$) within identical thin annular regions (Fig. \ref{fig2}a) and dividing by their respective areas. Finally, the areal density of permanent defects ($\psi_\mathrm{2}$) in the matrix after inclusion recovery was calculated as the sum of the defects produced during contraction ($\psi_\mathrm{1}$) and those produced during the recovery process ($\Delta\psi$). The choice of $\delta{r}=0.5L_\mathrm{f}$ for the annuli thickness was made to ensure a rigorous measure of spatial changes in fiber excess lengths; prior studies \cite{sarkar2022evolution,sarkar2024unexpected} and our results (Fig. \ref{fig3}f) revealed that excess lengths typically remain below sub-fiber length scales for individual fibers. For each measure of areal density of defects, $\psi_\mathrm{1}$, $\Delta\psi$, and $\psi_\mathrm{2}$, the respective summed excess lengths were non-dimensionalized by normalizing with the average matrix fiber length, $L_\mathrm{f}$.

To quantify the effective areal density of defects produced during the recovery process ($\Delta\psi_\mathrm{r}$) within a region of interest with radial dimension $h$ surrounding the inclusion, we calculated the average value of $\Delta\psi(r)$ between $r=0.25L_\mathrm{f}$ and $r=h$, as $\Delta\psi(r)$ typically remained positive beyond $r=0.25L_\mathrm{f}$ (Fig. \ref{fig3}c). To determine the total length of permanent defects ($\xi$) within the same region of interest, the areal density of permanent defects ($\psi_\mathrm{2}$) was integrated over both radial and angular increments to cover the entire area of the region. Conceptually, the total length of permanent defects ($\xi$) represent the sum of the permanent buckling-mediated defects in all fibers within the region of interest, quantified as the combined excess lengths of individual fibers during contraction and recovery processes, expressed as $\xi=\sum\left(\xi_\mathrm{f}+\Delta\xi_\mathrm{f}\right)$. As elaborated in the Results, normalized $\xi$ (non-dimensionalized by normalizing with $L_\mathrm{f}$) served as an internal variable, acting as a messenger of mechanical memory.

\subsubsection*{Measuring tension fibers}
Wherever discussed in the Results, the total length of fibers under axial tension within a designated region of interest in the matrix, denoted as $L_{\mathrm{eff}}$, was determined in accordance with established methodologies \cite{sarkar2022evolution,mann2019force}. This measurement was normalized by the total length of all fibers, $L_{\mathrm{t}}$, within the same region, to calculate the proportion of tension fibers.

\subsubsection*{Measuring microstructural gradient}
To quantify the alignment of matrix fibers during inclusion contraction, we measured an order parameter $S$, broadly inspired by the usage of similar parameters in describing the alignment of molecules in matter \cite{andrienko2018introduction}. The order parameter $S$ was measured at various distances $r$ from the inclusion within thin annuli of thickness $\delta{r} = 0.5L_\mathrm{f}$ at each $r$ (Fig. \ref{fig2}a). This method was similar to the measurement of defect evolution and aimed to assess the degree of radial alignment of each fiber relative to its reference orientation, averaging this alignment across all fibers within each annulus. The spatial gradient of $S$ was evaluated by calculating its first-order derivative with respect to the radial distance $r$ from the inclusion boundary and was expressed as $\left|dS/dr\right|$, where $\left|\right|$ indicates normalization of the absolute gradient measure by multiplying it with the average matrix fiber length $L_\mathrm{f}$. This choice of thin annuli ensured not only a rigorous measure of gradual spatial changes in fiber alignment but also consistency with the defect measurement approach. The gradient of fiber alignment, $\left|dS/dr\right|$, is referred to in this manuscript as the microstructural gradient.

The effective magnitude of the microstructural gradient $\left|dS/dr\right|_\mathrm{r}$ was measured in a manner similar to $\Delta\psi_\mathrm{r}$, within a region of interest with radial dimension $h$ surrounding the inclusion, by averaging the values of $\left|dS/dr\right|$ between $r=0.25L_\mathrm{f}$ and $r=h$.

\subsubsection*{Measuring matrix density and heterogeneity} 
Matrix density ($\rho$) was evaluated using the kernel density of material points, where denser clustering of material points indicated higher matrix density. This methodology is inspired by techniques described in Ref. \cite{vestal2021filtering}.  

To assess the extent to which permanent defects contributed to an increase in the geometric randomness of the remodeled matrix, we measured a heterogeneity parameter $\chi$, defined over a region of interest around the recovered inclusion. This parameter is conceptually similar to those employed in prior studies of fibrous matrices to characterize heterogeneous displacement fields \cite{head2003deformation,head2005mechanical,chandran2006affine,hatami2008scaling,grimmer2018displacement}. To quantify $\chi$, the Fourier transform of the absolute permanent displacement field of the matrix in the region of interest was performed to calculate the energy of displacement fluctuations across wave numbers. The sum of these energies across wave numbers represented the total spectral energy of the permanent displacement field. This quantification reflects the geometric heterogeneity $\chi$, capturing the additional geometric randomness imparted to the matrix by the processes of contraction and recovery of the inclusion. This method builds on the framework established in a previous study \cite{sarkar2022quantification}.  

\subsubsection*{Measuring matrix stiffness}
The incremental bulk stiffness ($k$) of the matrix was quantified after each load step as the change in the total internal resistance of the matrix to an incremental increase in fiber defects. Broadly inspired by Ref. \cite{parvez2023stiffening}, this stiffness was calculated as the second-order derivative of the total strain energy of all constituent fibers with respect to their total defects during the corresponding load step. This approach measured bulk stiffness rather than stiffness at a single point in the matrix, such as using a contracting dipole \cite{proestaki2021effect,sarkar2024unexpected}. Adopting a bulk measure of stiffness ensured that the inherent heterogeneity of the matrix did not obscure the measurement of stiffness, enabling a clearer understanding of the consequences of mechanical memory formation in fibrous materials.

\section*{Results}

To apply local tension on a matrix with free external boundaries, an embedded inclusion of diameter $D=2L_\mathrm{f}$ was contracted by varying radial strain $\epsilon_\mathrm{r}$ (Figs. \ref{fig1}, \ref{fig2}a). Following each contraction, the inclusion was allowed to recover by expanding back to its original size, releasing the applied tension within the matrix. The resulting mechanical states of the matrix surrounding both the contracted and recovered inclusion were analyzed. 

\subsection*{Contraction of inclusion generates local tension and defects}

\begin{figure}[t]
\centering
\includegraphics[width=6.5in, keepaspectratio=true]{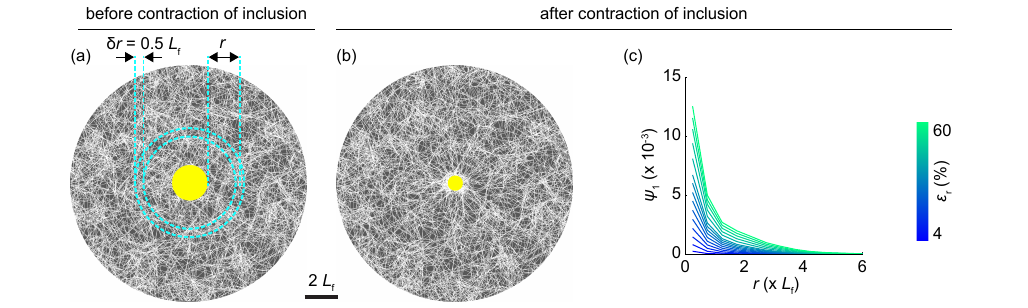}
\caption{Foundational insight as the inclusion contracts.
(a, b) Enlarged view of the representative matrix from Fig. \ref{fig1} before inclusion contraction (panel a) and after inclusion contraction by $60\%$ (panel b). Adjacent scale bar refers to panels a and b.
(c) Areal density of defects ($\psi_\mathrm{1}$) produced by the contracted inclusion as a function of distance $r$ from the inclusion boundary. $\psi_\mathrm{1}$ at each $r$ was measured over an annular region of thickness $\delta{r}=0.5L_\mathrm{f}$ (panel a).
Each line in panel c represents the average defect evolution across eight independent matrices, with the spread of data across these averages shown in Supplemental Fig. S2. 
Color bar indicates inclusion contraction levels ($\epsilon_\mathrm{r}$).
Background colors in panels a and b are to enhance graphic contrast of the matrix fibers.
}
\label{fig2}
\end{figure}

The contraction of the embedded inclusion, with radial strains ($\epsilon_\mathrm{r}$) ranging from $4\%$ to $60\%$, induced local tension on the surrounding matrix fibers (Fig. \ref{fig2}b, \ref{fig3}a). Echoing observations from a previous study \cite{burkel2017mechanical}, this tension led to the formation of two visually distinct fiber populations (Figs. \ref{fig2}b): one set aligned radially and the other circumferentially, parallel to the periphery of the inclusion, with the circumferential fibers potentially buckling. Buckled circumferential fibers introduced mechanical instabilities, defined as defects in this study, which in turn promoted the alignment of radial fibers.

The spatial evolution of defects, characterized by their areal density ($\psi_\mathrm{1}$) measured at various distances $r$ from the inclusion boundary, revealed that defects originated at the boundary of the contracted inclusion and propagated outward, extending up to several fiber lengths into the matrix (Fig. \ref{fig2}c). This defect propagation is a known manifestation of buckling-driven, long-range transmission of deformation in fibrous matrices \cite{rudnicki2013nonlinear, burkel2017mechanical}. Importantly, higher radial contraction resulted in an increase in defect formation near the inclusion (Fig. \ref{fig2}c). These defects, as prior studies \cite{grekas2021cells, proestaki2019modulus} depicted, drive nonlinear mechanical responses within the matrix microenvironment of the contracted inclusion; see Supplemental Fig. S1 and Supplemental Note 2 for details on how our model reproduced these nonlinear responses.

\subsection*{Recovering inclusion produces defects rather than eliminating them}

\begin{figure}[ht!]
\centering
\includegraphics[width=6.5in, keepaspectratio=true]{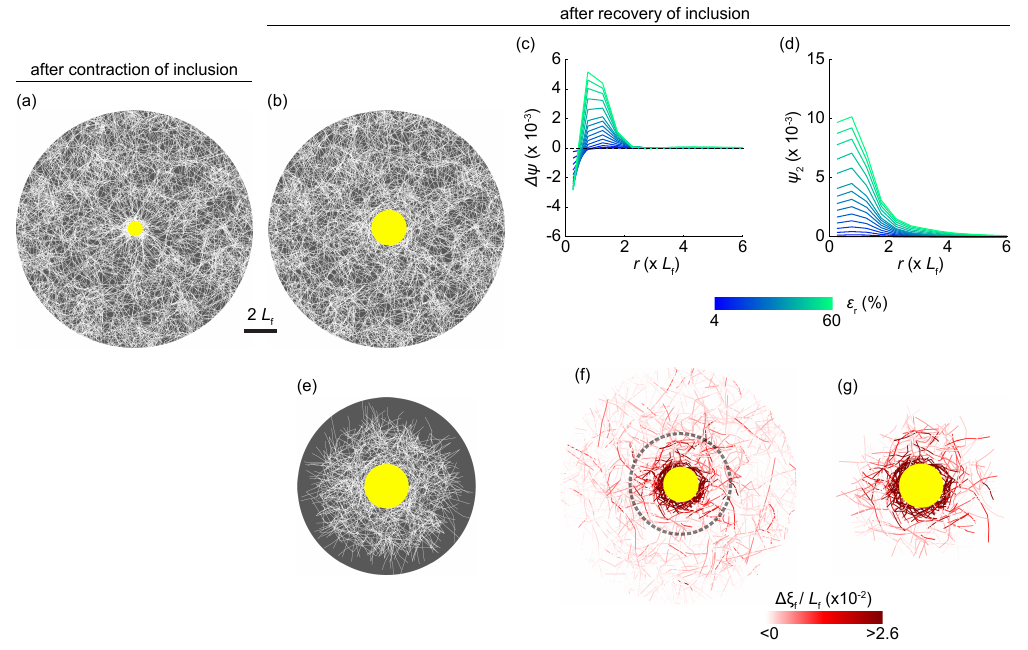}
\caption{Recovery of inclusion produces defects. 
(a) Enlarged view of the representative matrix from Fig. \ref{fig1} after inclusion contracted by $60\%$. 
(b) Matrix from panel a after the contracted inclusion recovered. Scale bar refers to panels a and b.
(c) Areal density of defects ($\Delta\psi$) produced by the recovery process, as a function of distance $r$ from the inclusion boundary.  
(d) Areal density of permanent defects ($\psi_\mathrm{2}$), as a function of distance $r$ from the inclusion boundary. $\psi_\mathrm{2}$ represents the sum of defects produced during the contraction of the inclusion ($\psi_\mathrm{1}$, Fig. \ref{fig2}c) and the defects produced by the recovery process ($\Delta\psi$, panel c). 
For panels c and d, quantities at each $r$ were measured over an annular region of thickness $\delta{r}=0.5L_\mathrm{f}$, as depicted in Fig. \ref{fig2}a.
Color bar indicates inclusion contraction levels ($\epsilon_\mathrm{r}$) prior to recovery for lines in panels c and d.
(e) A representative region of interest spanning $2L_\mathrm{f}$ around the recovered inclusion where defect production peaked during the recovery process. 
(f, g) In the representative matrix after inclusion recovery (where the inclusion was contracted by $60\%$ before recovery), fibers induced with defects by the recovery process ($\Delta\xi_\mathrm{f} > 0$) are shown in progressive red, with color intensity reflecting the extent of defects induced. Panel f depicts fibers around the recovered inclusion, and panel g zooms into the $2L_\mathrm{f}$ region of interest (encircled in panel f). 
Values of $\Delta\xi_\mathrm{f}$ are normalized by $L_\mathrm{f}$.
In panels e--g, the diameter of the recovered inclusion, $2L_\mathrm{f}$, serves as the scale bar. 
Background colors in panels a, b, and e are employed to enhance graphic contrast. Each line in panels c and d represents the average defect evolution across eight independent matrices, with the spread of data across these means shown in Supplemental Fig. S2.
}
\label{fig3}
\end{figure}

Following contraction, a recovery process ensued as the inclusion expanded back to its original size (Fig. \ref{fig3}b). The conventional notion of elasticity, following prior studies \cite{kim2017stress, ban2018mechanisms, grekas2021cells}, predicts that buckling-mediated defects generated during the inclusion contraction would be reversed during the recovery process, restoring the matrix fibers to their reference state. Contrary to these expectations, our observations revealed a surprising outcome: the recovery process produced new defects ($\Delta\psi$) in the surrounding matrix (Fig. \ref{fig3}c). These defects, produced during recovery ($\Delta\psi$), added to those initially generated during the contraction of the inclusion ($\psi_\mathrm{1}$, Fig. \ref{fig2}c), collectively resulted in permanent defects ($\psi_\mathrm{2}$, Fig. \ref{fig3}d), signifying substantial mechanical remodeling of the matrix.

The recovery-induced defects ($\Delta\psi$) evolved with distance from the inclusion boundary (Fig. \ref{fig3}c), showing that greater inclusion contraction prior to recovery resulted in an increase in defect formation during recovery and, consequently, higher levels of permanent defects ($\psi_\mathrm{2}$, Fig. \ref{fig3}d). Additionally, the recovery process consistently produced the majority of defects ($\Delta\psi$) within $\approx2L_\mathrm{f}$ from the inclusion boundary, regardless of the level of inclusion contraction before recovery (Fig. \ref{fig3}c,e). To further support this assertion, a fiber-level map (Fig. \ref{fig3}f) illustrating the extent of defects induced in fibers by the recovery process ($\Delta\xi_\mathrm{f}$) indicates a higher number of fibers with pronounced defects concentrated within this $2L_\mathrm{f}$ region around the inclusion (Fig. \ref{fig3}g, Supplemental Fig. S2). These observations prompt the need to investigate the mechanisms likely activated during the recovery process, specifically within the vicinity spanning $2L_\mathrm{f}$ from the inclusion boundary.

\subsection*{Inclusion recovery amidst microstructural gradient}

\begin{figure}[t]
\centering
\includegraphics[width=6.5in, keepaspectratio=true]{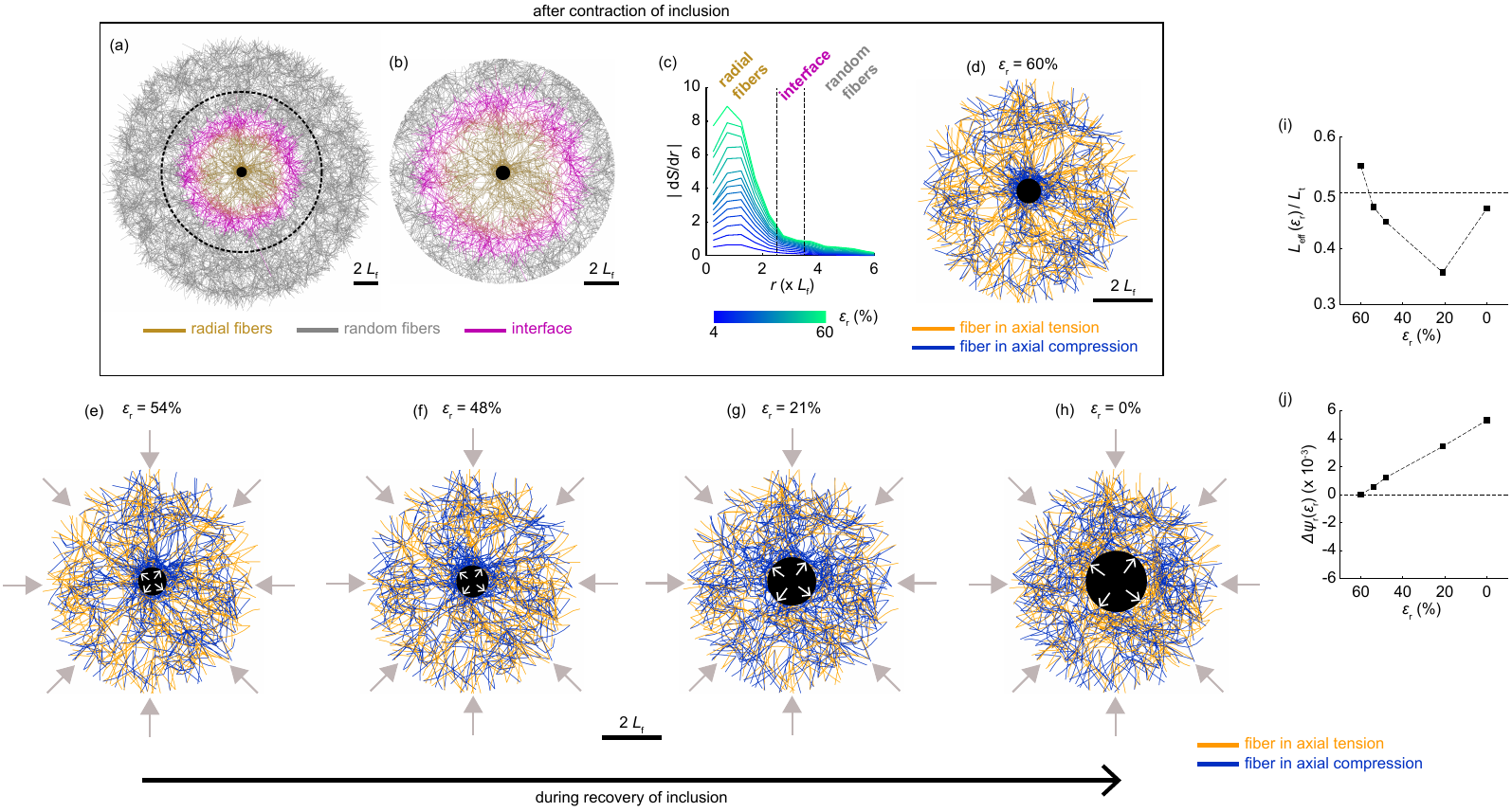}
\caption{Microstructural gradient and recovery-induced defects.
(a,b) Matrix fibers near the $60\%$ contracted inclusion show prominence in alignment within the yellow region and an absence of alignment in the grey region, separated by an interface highlighted in magenta. Panel b zooms into the encircled region in panel a. 
(c) The evolution of the microstructural gradient, $\left|{dS}/{dr}\right|$, with distance $r$ from the inclusion boundary. Color bar indicates inclusion contraction levels ($\epsilon_\mathrm{r}$).
(d) Matrix region of interest within $2L_\mathrm{f}$ from the contracted inclusion, color-coded to show fibers in axial compression (blue) or tension (orange), with a high proportion of fibers in tension.
(e--h) The matrix region of interest (panel d) as the inclusion recovers to its original size, with grey arrows external to matrix schematically indicating radial confinement by the microstructural gradient, creating radial compression. 
(i, j) Evolution of the proportion of tension fibers ($L_\mathrm{eff}(\epsilon_\mathrm{r})/L_\mathrm{t}$) and the production of defects ($\Delta\psi_\mathrm{r}(\epsilon_\mathrm{r})$) as the inclusion recovers, for matrix states depicted in panels e--h. 
$\Delta\psi_\mathrm{r}$ is the effective defect density within the $2L_\mathrm{f}$ region of interest as described in the Methods.
Each line in panel c represents the average of eight independent matrices. 
Data spread across these averages for panel c is available in Supplemental Fig. S3.
}
\label{fig4}
\end{figure}

To elucidate the mechanisms by which the recovering inclusion generated defects, we analyzed the evolving microstructure of the matrix throughout the loading history, both during the initial contraction and the subsequent recovery of the inclusion. Initially, the matrix fibers were randomly oriented, but upon contraction of the inclusion, the surrounding matrix organized into two distinct zones (Fig. \ref{fig4}a,b, Supplemental Fig. S3). In the immediate vicinity of the contracted inclusion, a significant proportion of fibers aligned radially with the inclusion (highlighted in yellow). Beyond this, a transitional interface emerged where fiber alignment gradually shifted from radial to random (highlighted in magenta), followed by a second zone where fibers remained predominantly disordered (highlighted in grey). Quantitative analysis revealed these spatial changes in fiber alignment by measuring the order parameter $S$, which depicts fiber alignment, at increasing distances $r$ from the inclusion (see Methods and Supplemental Fig. S2) and computing its spatial gradient, $\left|dS/dr\right|$ (Fig. \ref{fig4}c). As this alignment gradient, $\left|dS/dr\right|$, illustrated the strain-induced changes in the matrix microstructure resulting from the radial contraction of the inclusion, it was referred to as the microstructural gradient for simplicity throughout this manuscript.

While the microstructural gradient was greater at a higher level of inclusion contraction ($\epsilon_\mathrm{r}$), it typically peaked within $r<2L_\mathrm{f}$ (Fig. \ref{fig4}c). An interface was identified between $r=2L_\mathrm{f}$ and $r=4L_\mathrm{f}$, where the gradient gradually diminished, and beyond $r>4L_\mathrm{f}$, no gradient was observed (Fig. \ref{fig4}c). Notably, the localization of both the microstructural gradient and the earlier identified recovery-induced defects ($\Delta\psi$, Fig. \ref{fig3}c) within $2L_\mathrm{f}$ from the inclusion boundary indicates that this region spanning $2L_\mathrm{f}$ around the inclusion (Fig. \ref{fig4}d) should be the region of interest for further investigation into the mechanisms underlying defect formation. The strain-induced microstructural gradient within the matrix, reminiscent of similar gradients in granular materials that restrict structural mobility \cite{keim2019memory,paulsen2024mechanical}, suggests potential obstructions to fiber movement and changes in local stress distributions during inclusion recovery.

To investigate the consequences of the strain-induced microstructural gradient during the recovery of the inclusion (Fig. \ref{fig4}d--h), which contracted by $60\%$ in this representative case, we analyzed the stress states of fibers within the region of interest spanning $2L_\mathrm{f}$ from the inclusion boundary. At the onset of recovery (Fig. \ref{fig4}d), the fraction of radial tension fibers (highlighted in orange) exceeded that of circumferential compression fibers (highlighted in blue), with $55\%$ of the fibers bearing axial tension. We then monitored the evolution of fiber stress states in the region of interest as the contracted inclusion expanded toward its recovery (Fig. \ref{fig4}e--h). Although the expanding boundary of the inclusion sought to release existing stresses on both radial tension fibers and circumferential compression fibers, the microstructural gradient imposed geometric confinement, schematically depicted by grey radial arrows surrounding the region of interest in Fig. \ref{fig4}e--h. This confinement restricted the outward radial movement of fibers near the expanding inclusion, preventing them from returning to the configuration they had before the inclusion was contracted. 

The radial confinement imposed by the microstructural gradient particularly affected the radially aligned fibers, inhibiting the recovery of their stress states and leading to a build-up of axial compression and potential buckling. This effect was reflected in the increasing proportion of compression fibers as the inclusion continued to recover (Fig. \ref{fig4}d--g), with the fraction of tension fibers gradually decreasing from $55\%$ at $\epsilon_\mathrm{r}=60\%$ to $35\%$ at $\epsilon_\mathrm{r}=21\%$ in the region of interest (Fig. \ref{fig4}i). The progressive fiber buckling that was initiated near the recovering inclusion (Fig. \ref{fig4}e--g, Supplemental Fig. S4), produced new defects ($\Delta\psi_\mathrm{r} (\epsilon_\mathrm{r})>0$, Fig. \ref{fig4}j). Intriguingly, as the inclusion continued to expand further from $\epsilon_\mathrm{r}=21\%$ to its fully recovered state at $\epsilon_\mathrm{r}=0\%$ (Fig. \ref{fig4}h), tension reappeared in fibers near the inclusion boundary, increasing the fraction of tension fibers from $35\%$ to $47\%$. This resurgence of tension resulted from severe post-buckling distortions in already buckled fibers, as evidenced by the increase in defects between matrix states at $\epsilon_\mathrm{r}=21\%$ and $\epsilon_\mathrm{r}=0\%$ (Fig. \ref{fig4}j) that are driven by the advancing boundary of the recovering inclusion. Therefore, the geometric confinement imposed by the microstructural gradient led to both buckling and post-buckling distortions in fibers near the recovering inclusion, implying a gradual production of defects during the recovery process.

\begin{figure}[hbt!]
\centering
\includegraphics[width=3.3in, keepaspectratio=true]{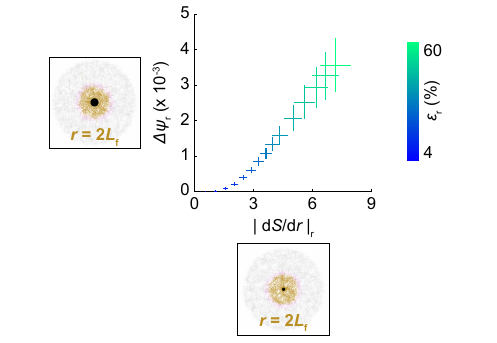}
\caption{Novel history-dependent behavior in the matrix as microstructural gradient ($\left|{dS}/{dr}\right|_\mathrm{r}$) influences defect production ($\Delta\psi_\mathrm{r}$) during the recovery of inclusion. 
Both $\left|{dS}/{dr}\right|_\mathrm{r}$ and $\Delta\psi_\mathrm{r}$ represent the effective magnitudes of the gradient and defect density, respectively, within the region of interest $2L_\mathrm{f}$ around the inclusion (Fig. \ref{fig4}d); see Methods for details on their evaluation based on spatial evolution.
Each data point represents the average from eight independent matrices, with error bars indicating standard error. 
Color bar indicates inclusion contraction levels ($\epsilon_\mathrm{r}$) prior to recovery.
Insets along the axes illustrate a representative case of inclusion contraction: the lower inset highlights the region of interest around the contracted inclusion where $\left|{dS}/{dr}\right|_\mathrm{r}$ was calculated, and the side inset shows the same region around the recovered inclusion where $\Delta\psi_\mathrm{r}$ was assessed.
}
\label{fig5}
\end{figure}

In our analysis of a representative matrix in which the inclusion had contracted by $60\%$ (Fig. \ref{fig4}d--j), we observed that the microstructural gradient initiated by this contraction facilitated the production of defects during the recovery process. Further quantitative analysis helped to understand how the extent of inclusion contraction influenced both the microstructural gradient and defect production within the region spanning $2L_\mathrm{f}$ around the inclusion. Notably, greater levels of inclusion contraction led to higher microstructural gradients ($\left|{dS}/{dr}\right|_\mathrm{r}$) and increased defect production during recovery ($\Delta\psi_\mathrm{r}$) (Fig. \ref{fig5}). Thus, inclusion contraction, via the microstructural gradient, uniquely tuned defect production in the matrix during recovery, embedding a mechanical memory that persisted even after the inclusion had recovered—a finding that reveals a novel history-dependent behavior in fibrous matrices.

\subsection*{Defects tune matrix properties around recovered inclusion}

\begin{figure}[t!]
\centering
\includegraphics[width=6.5in, keepaspectratio=true]{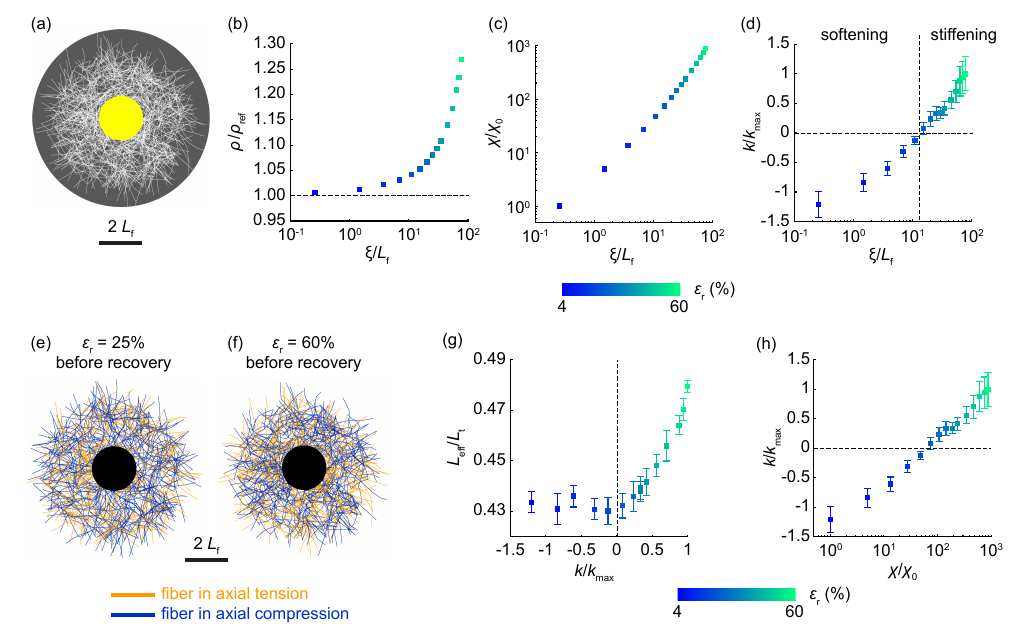}
\caption{Defects tune matrix properties in the region of interest. 
(a) Representative region of interest spanning $2L_\mathrm{f}$ around the recovered inclusion. Here, the inclusion contracted by $60\%$ before recovery. 
(b, c) Higher inclusion contraction ($\epsilon_\mathrm{r}$) led to greater extent of permanent defects ($\xi$) and higher densification ($\rho$) (panel b), and added more geometric heterogeneity ($\chi$) (panel c). 
Density $\rho$ is normalized by the density of the reference (undeformed) matrix ($\rho_\mathrm{ref}$), and heterogeneity $\chi$ is normalized by the heterogeneity at the lowest $\epsilon_\mathrm{r}=4\%$ ($\chi_\mathrm{0}$). 
(d) Incremental bulk stiffness ($k$) of the matrix, showing either softening or stiffening around $\epsilon_\mathrm{r}\approx21\%$. Stiffness is normalized against the maximal mean stiffness ($k_\mathrm{max}$) of the independent matrix models considered in this study. 
In panels b--d, $\xi$ was non-dimensionalized by the average matrix fiber length $L_\mathrm{f}$.
(e, f) Representative matrix around the recovered inclusion showing fibers under axial tension (orange) and compression (blue) at different values of $\epsilon_\mathrm{r}$ prior to recovery. 
(g) Stiffening ($k > 0$) occurred when the proportion of tension fibers ($L_\mathrm{eff}/L_\mathrm{t}$) increased.  
(h) Escalating matrix heterogeneity (from panel c) boosts stiffness (from panel d).
Panels b–d, g, and h relate to properties within the matrix region shown in panel a. Each data point in panels b–d, g, and h corresponds to means from eight independent matrices. Error bars in panels d, g, and h show standard errors. For the full spread of data at each data point in panels b–d, g, and h, see Supplemental Fig. S5. Color bars adjacent to panels b–d, g, and h indicate inclusion contraction levels ($\epsilon_\mathrm{r}$) for data points.
}
\label{fig6}
\end{figure}

In the previous sections, we documented how the processes of contraction and subsequent recovery of an inclusion individually contributed to the production of permanent defects within the matrix (Fig. \ref{fig3}d). Furthermore, the spatial evolution of these defects, driven by geometric gradients, revealed an assemblage of defects altering fiber stress states near the inclusion boundary within a defined region of interest spanning $2L_\mathrm{f}$ (Fig. \ref{fig4}). This observation prompted further investigation into how matrix properties in this region varied based on the extent of permanent defects present. To address this, we shifted our focus from areal density to a total measure of permanent defects ($\xi$, see Methods) within the fixed $2L_\mathrm{f}$ area around the inclusion (Fig. \ref{fig6}a). This measure of the total length of permanent defects ($\xi$) retained a record of the kinematic changes in the matrix fibers pursuant to its loading history, specifically the levels of inclusion contraction prior to recovery, establishing the total length of permanent defects as an internal variable that captured the memory of the loading history and defined the post-recovery state of the matrix.

Permanent defects ($\xi$) rendered the matrix more dense ($\rho$, Fig. \ref{fig6}b), with the density being greater at a higher extent of defects and a higher level of inclusion contraction ($\epsilon_\mathrm{r}$). In the matrix we studied, which initially exhibited a disordered, random microstructure, the accumulation of permanent defects that are mediated by buckled and distorted fibers, introduced an additional layer of geometric randomness, or heterogeneity. As described in the Methods, this induced geometric heterogeneity was quantified using the parameter $\chi$, which, akin to matrix density, increased at higher $\xi$ and $\epsilon_\mathrm{r}$ (Fig. \ref{fig6}c). Thus, the permanent defects fundamentally altered the physical state of the matrix, enhancing both its density and heterogeneity.

Next, we examined how permanent defects influenced the bulk stiffness ($k$) of the matrix surrounding the recovered inclusion. To this end, we adopted an incremental measure of stiffness, as described in Methods, aligning with practices in architected materials design (for example, Ref. \cite{liu2021origin}), focusing our analysis on how changes in defect production ($\xi$) and fiber stress states modulated matrix response in the region of interest. Depending on the extent of inclusion contraction prior to recovery, the matrix exhibited either softening or stiffening, reflected by a negative or positive value of incremental stiffness, respectively, with stiffening typically observed when $\epsilon_\mathrm{r} > 21\%$ (Fig. \ref{fig6}d).

The explanation for these stiffness variations is rooted in the stress state of the matrix. At force equilibrium, the matrix surrounding the recovered inclusion remained under radial compression due to the outward push of the expanded boundary of the inclusion and the confining pressure from the microstructural gradient (Fig. \ref{fig4}e--g). At lower to moderate levels of inclusion contraction ($\epsilon_\mathrm{r}\le21\%$), gradient-driven confinement facilitated fiber buckling as the inclusion recovered, softening the matrix—a phenomenon somewhat analogous to softening observed in fibrous matrices under low to moderate compression \cite{vahabi2016elasticity, van2016uncoupling, notbohm2015microbuckling, rosakis2015model, kim2014structural}. Conversely, higher levels of inclusion contraction intensified microstructural gradients, enhancing radial compression and stiffening the matrix, similar to known compression-stiffening effects observed in fibrous matrices under high compression \cite{shivers2020compression}. This stiffening was induced by severe post-buckling distortions in fibers near the boundary of the recovered inclusion, with some fibers transitioning to tension (Figs. \ref{fig4}h, \ref{fig6}e,f). Further quantification showed that as contraction exceeded $\epsilon_\mathrm{r}>21\%$, the proportion of tension fibers ($L_\mathrm{eff}/L_\mathrm{t}$) increased, elevating matrix stiffness (Fig. \ref{fig6}g).

Based on the premise that the size of the matrix affects its mechanics \cite{shahsavari2013size, tyznik2019length, merson2020size}, we evaluated how the thickness of the annulus region surrounding the recovered inclusion influenced stiffness by studying annuli of thickness $h = 2$, $1$, and $0.5 L_\mathrm{f}$ (Supplemental Fig. S6a--c). We observed consistent trends in mechanical responses, with the matrix either softening or stiffening depending on the levels of permanent defects produced (Supplemental Fig. S6d). Notably, instances of stiffening were consistently associated with a higher fraction of tension fibers (Supplemental Fig. S6e).

In summary, the greater the permanent defects ($\xi$), the more pronounced were the increases in both heterogeneity ($\chi$, Fig. \ref{fig6}c) and stiffness ($k$, Fig. \ref{fig6}d), suggesting that escalating matrix heterogeneity is a crucial driver of the observed increase in stiffness (Fig. \ref{fig6}h). This phenomenon resonates with findings from a previous study \cite{shivers2020compression} that explored matrix behavior within the compression-stiffening regime. 

Overall, our findings revealed that an applied tensile deformation imprints a unique geometric signature in the matrix by establishing a distinct microstructural gradient, which, upon the release of tension, sets permanent defects as an internal variable that defines the mechanical state of the matrix, illustrating an intrinsic property of the matrix to retain a memory of its deformation history.

\section*{Discussion}
Our numerical studies demonstrated that the localized tensile boundary condition applied by an inclusion during contraction, upon release, caused radial compression in the surrounding matrix. This matrix remodeling was surprising, as our models did not include explicit plasticity mechanisms such as inter-fiber cohesion or fiber yielding, thereby challenging the traditional assumption of elasticity in mechanics models of fibrous matrices. Moreover, the resulting matrix remodeling, characterized by changes in density and stiffness, reflected the memory of past deformation. We demonstrated that a specific level of applied tension imprints this memory in the fibrous matrix by creating a distinct magnitude of microstructural gradient, which in turn obstructs fiber movement during the release of tension and drives the remodeling process. Unraveling this memory mechanism in disordered fibrous matrices, governed entirely by geometry and prior mechanical loading, is a novel contribution of this manuscript.

In this study, the total length of permanent defects ($\xi$) functioned as an internal variable that preserved the memory of past tensile boundary conditions in the matrix. Introducing these defects as an internal variable provided a novel perspective, especially when contrasted with traditional material characterization methods in classical plasticity theory, where internal variables typically describe residual strains \cite{rice1971inelastic}. This concept of an internal variable, particularly in the context of fibrous matrices, extends beyond the classical description of strain to include features such as fiber rotation \cite{favata2022internal}. Building on this concept, given the inevitability of mechanical instabilities leading to persistent buckling-mediated defects, we adopted the total length of permanent defects as a broadly applicable internal variable. This metric, derived from the sum of the buckling extent across all fibers within the region of interest, underscores the pivotal role of fiber buckling in preserving mechanical memory in fibrous matrices. This extends a decade of research underscoring the impact of fiber buckling on various mechanical phenomena within these materials. Buckling introduces mechanical instabilities at the fiber scale, contributing to local densification \cite{grekas2021cells,proestaki2022effect} and the long-range propagation of mechanical cues \cite{doha2022disorder,grimmer2018displacement,rosakis2015model,notbohm2015microbuckling,goren2020elastic}. Previous studies have documented fiber buckling under conditions such as global shear \cite{munster2013strain}, global compression \cite{kim2014structural, shivers2020compression}, and localized tension \cite{burkel2017mechanical,proestaki2019modulus,sarkar2024unexpected}. This study advances our understanding by identifying a new role of fiber buckling as a history-dependent internal variable that preserves the loading history of the matrix, offering nuanced insights into how disordered fibrous materials retain and utilize mechanical memory.

Distinctive aspects of our model stand out, especially when compared with recent models of fibrous materials. Unlike the approach in Refs. \cite{grekas2021cells,ban2018mechanisms}, we did not consider fiber adhesion essential for sustaining matrix deformation after the recovery of the inclusion. Although our observations resonated with a recent continuum model \cite{favata2024emerging}, which suggested that fibrous materials might inherently retain loading history through geometry alone, our study expanded on this concept by uncovering an additional mechanism at play. Specifically, we identified gradient-induced fiber buckling as an internal variable encoding mechanical memory independent of traditional plasticity.

Interestingly, the strain-induced gradient observed in fibrous matrices also draws parallels to the microstructural gradients engineered in alloys, where such gradients restrict the movement of structural features like dislocations and alter local stress fields \cite{duan2024order, li2020mechanical, wu2021extraordinary, shao2018simultaneous, hughes2018microstructural, ding2018mechanical}. Specifically, the contraction of an inclusion in fibrous matrices generated a microstructural gradient that formed an ordered core of aligned fibers surrounded by a disordered shell of randomly oriented fibers (Fig. \ref{fig4}b), resembling the core-shell structures recently developed in alloys \cite{duan2024order}. This ordered core encased by a disordered shell has been shown to enhance mechanical properties by promoting dislocation accumulation under cyclic loading. Similarly, in fibrous matrices, strain-induced gradients contribute to defect formation within the ordered core near the inclusion, potentially increasing mechanical stiffness when a significant extent of defects is produced (Fig. \ref{fig6}d).

Our findings raise new questions and pave the way for further research. First, it is coincidental that the observed defects were contained within $2L_\mathrm{f}$ around the recovered inclusion (Fig. \ref{fig3}c), matching the undeformed diameter of the inclusion. While this study does not aim to establish a direct link between the inclusion diameter and the spatial extent of matrix remodeling, future research could explore this relationship. Second, the disparate effects observed from remodeling by multiple contracted inclusions versus a single one \cite{burkel2017mechanical, sarkar2024unexpected} set the stage for future studies on matrix responses to simultaneously recovering inclusions. Third, more broadly, our findings suggest that the ability to manipulate mechanical stiffness through strain-mediated microstructural gradients could refine existing design criteria for disordered fibrous metamaterials \cite{huang2023jammed, rayneau2019density}, heralding a transformative approach in material design that aims to create a new class of metamaterials with promising applications from flexible electronics to biomedical implants \cite{khalid20243d}. Overall, our study not only advances the understanding of the mechanics of fibrous matrices but also establishes a strategic foundation for the next frontier in materials design.

\section*{Acknowledgments}
MS and CL acknowledge the support from Charles Bell, Catherine Bracken and Barbara Bell. AWJ is a Chan Zuckerberg Biohub Chicago Investigator.

\section*{Declarations}
\subsection*{Conflicts of Interest}
The authors have no conflicts of interest to declare.

\bibliographystyle{elsarticle-num}
\bibliography{references.bib}

\newpage

\begin{center}

\textbf{Supplemental Information}

\Large
\textbf{Geometry-Driven Mechanical Memory in a Random Fibrous Matrix}
\vspace{11pt}

Mainak Sarkar,$^\mathrm{a,*}$ Christina Laukaitis,$^\mathrm{a,c,e,g}$ Amy Wagoner Johnson$^\mathrm{a,b,c,d,f}$
\vspace{11pt}

\small
$^\mathrm{a}$Carl R. Woese Institute for Genomic Biology, University of Illinois Urbana-Champaign \\ $^\mathrm{b}$ Mechanical Science and Engineering, Grainger College of Engineering, University of Illinois Urbana-Champaign \\ $^\mathrm{c}$Biomedical and Translational Sciences, Carle Illinois College of Medicine, University of Illinois Urbana-Champaign \\  $^\mathrm{d}$Beckman Institute for Advanced Science and Technology, University of Illinois Urbana-Champaign \\ $^\mathrm{e}$Clinical Sciences, Carle Illinois College of Medicine, University of Illinois Urbana-Champaign \\ $^\mathrm{f}$CZ Biohub Chicago, LLC, Chicago, Illinois \\ $^\mathrm{g}$ Carle Health, Urbana, Illinois \\ $^\mathrm{*}$ Corresponding author: Mainak Sarkar, mainak@illinois.edu \\

\end{center}

\section*{Supplemental Figures}



\begin{center}
\includegraphics[width=6.5in, keepaspectratio=true]{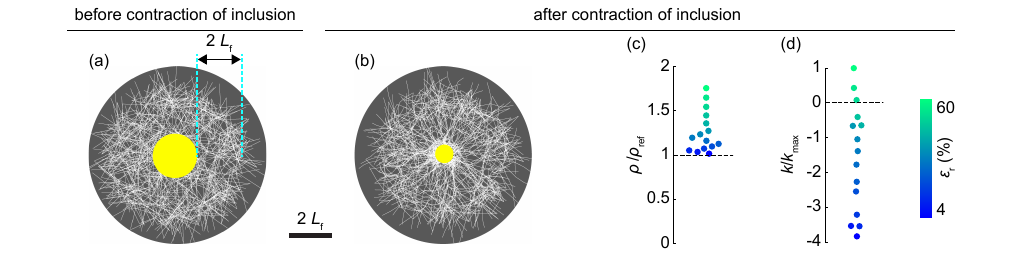}
\end{center}
\vspace{-18pt plus 2pt minus 2pt}
\textbf{Figure S1.} Contracted inclusion drives nonlinear mechanical responses.
(a, b) Representative region of interest spanning $2L_\mathrm{f}$ around the inclusion, shown both before (panel a) and after (panel b) inclusion contraction by $60\%$.
(c) Matrix density ($\rho$), normalized by the reference configuration density ($\rho_\mathrm{ref}$), within the region of interest as the inclusion undergoes radial contractions ranging from $4\%$ to $60\%$.
(d) Incremental bulk stiffness ($k$) of the matrix within the region of interest as the inclusion experiences radial contractions ranging from $4\%$ to $60\%$, normalized by the maximal stiffness ($k_\mathrm{max}$) observed at the highest inclusion contraction.
Data points in panels c and d are means from eight independent matrices. 
Background color in panels a and b are to enhance graphic contrast.
Color bar indicates inclusion contraction levels ($\epsilon_\mathrm{r}$) for data points in panels c and d.


\begin{center}
\includegraphics[width=6.5in, keepaspectratio=true]{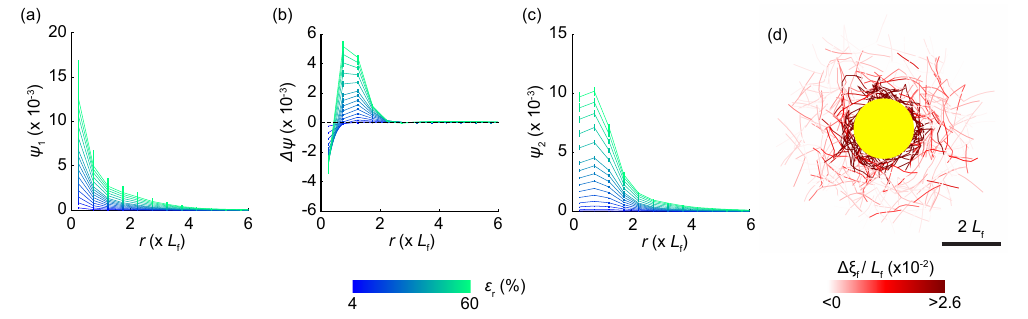}
\end{center}
\vspace{-14pt plus 2pt minus 2pt}
\textbf{Figure S2.}
Production of defects. 
(a, b) Areal density of defects produced by the process of inclusion contraction ($\psi_\mathrm{1}$, panel a) and inclusion recovery ($\Delta\psi$, panel b), showing the standard errors of data points across means taken from eight independent matrices. 
(c) Sum of the areal densities of defects produced during the process of inclusion contraction (panel a) and recovery (panel b), depicting the permanent defects ($\psi_\mathrm{2}$) in the matrix sustained after the inclusion has recovered. 
This panel also shows the standard errors of data points across means taken from eight independent matrices. 
Color bar indicates inclusion contraction levels ($\epsilon_\mathrm{r}$) for lines in panels a–c. 
(d) For another representative matrix different from that shown in Fig. 3, recovery process-induced defects in fibers within $2L_\mathrm{f}$ around the inclusion are depicted, with the extent of defects shown in progressive shades of red.

\begin{center}
\includegraphics[width=6.5in, keepaspectratio=true]{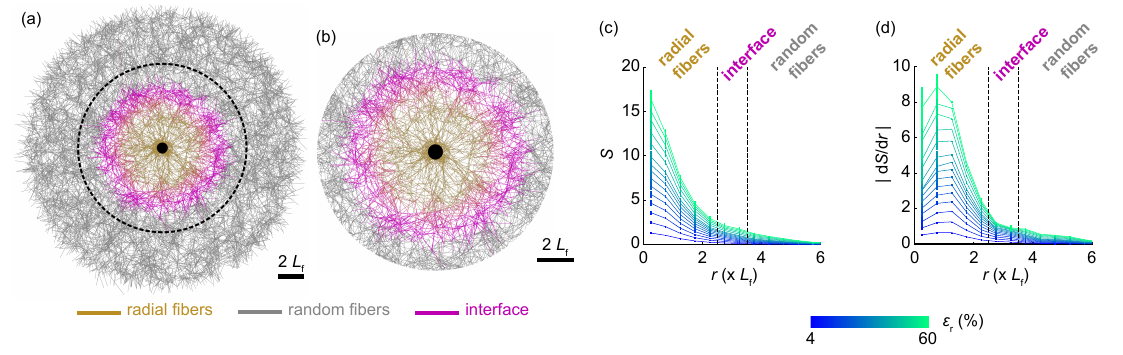}
\end{center}
\vspace{-16pt plus 2pt minus 2pt}
\textbf{Figure S3.}
Derivation of gradient microstructure.
(a, b) Another representative matrix (different from that shown in Fig. 4) around a contracted inclusion (inclusion contracted by $60\%$) showing a region of radially aligned fibers (yellow) surrounded by random fibers (grey), separated by an interface region (magenta). Panel b zooms into the encircled portion in panel a.
(c) The order parameter $S$ measures the radial alignment of matrix fibers, evolving with distance ($r$) from the contracted inclusion boundary, indicating elevated alignment (high $S$) near the inclusion that decreases with distance.
(d) The evolution of the microstructural gradient, $|{dS}/{dr}|$, with distance $r$ from the inclusion boundary.
Panels c and d depict standard errors across means taken from eight independent matrices.
Color bar indicates inclusion contraction levels ($\epsilon_\mathrm{r}$) for lines in panels c and d.
 
\begin{center}
\includegraphics[width=6.5in, keepaspectratio=true]{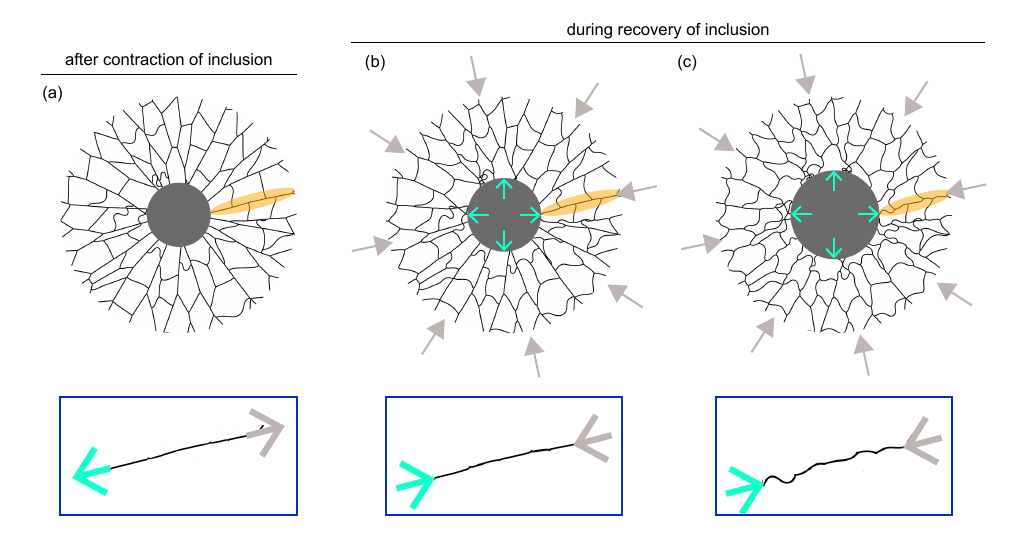}
\end{center}
\vspace{-11pt plus 2pt minus 2pt}
\textbf{Figure S4.}
Cartoon schematic of matrix with inclusion depicting confinement due to gradient microstructure.
(a) Matrix with a contracted inclusion. Inset shows a radially aligned fiber under axial tension.
(b, c) As the contracted inclusion expands during recovery, surrounding matrix fibers are constrained to move radially outward, confined by the gradient microstructure (indicated by grey arrows). This confinement progressively induces axial compression in the matrix fibers (panel b and inset), leading to eventual buckling (panel c and inset) or, in some cases, severe post-buckling (not shown), beginning near the inclusion boundary.


\begin{center}
\includegraphics[width=6.5in, keepaspectratio=true]{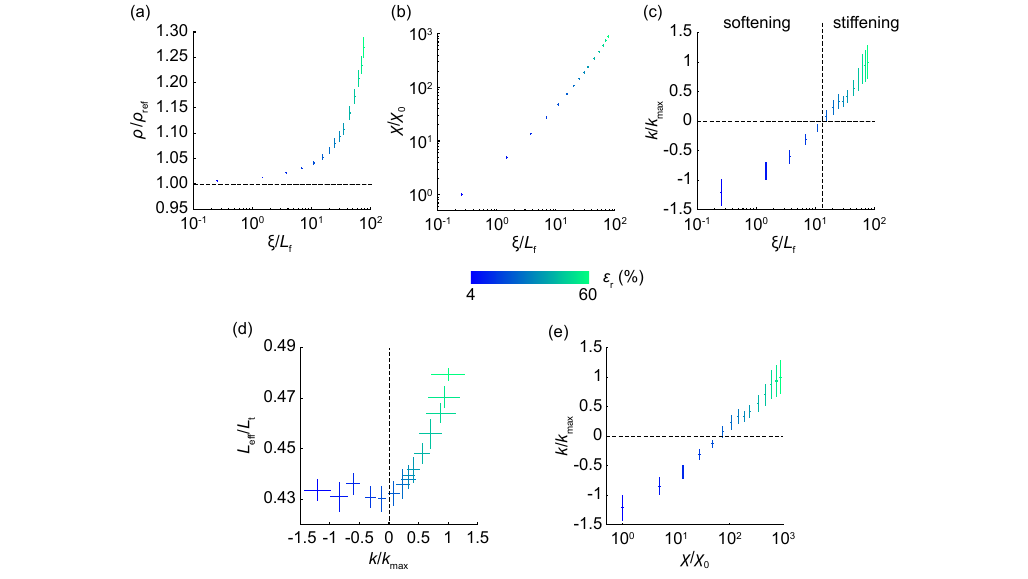}
\end{center}
\vspace{-18pt plus 2pt minus 2pt}
\textbf{Figure S5.}
Standard error representation for data points across means from eight independent samples, for plots in Fig. 6: (a) Standard errors for data in Fig. 6b. (b) Standard errors for data in Fig. 6c. (c) Standard errors for data in Fig. 6d. (d) Standard errors for data in Fig. 6g. (e) Standard errors for data in Fig. 6h.
Color bar indicates inclusion contraction levels ($\epsilon_\mathrm{r}$) for data points in all panels.

\begin{center}
\includegraphics[width=6.5in, keepaspectratio=true]{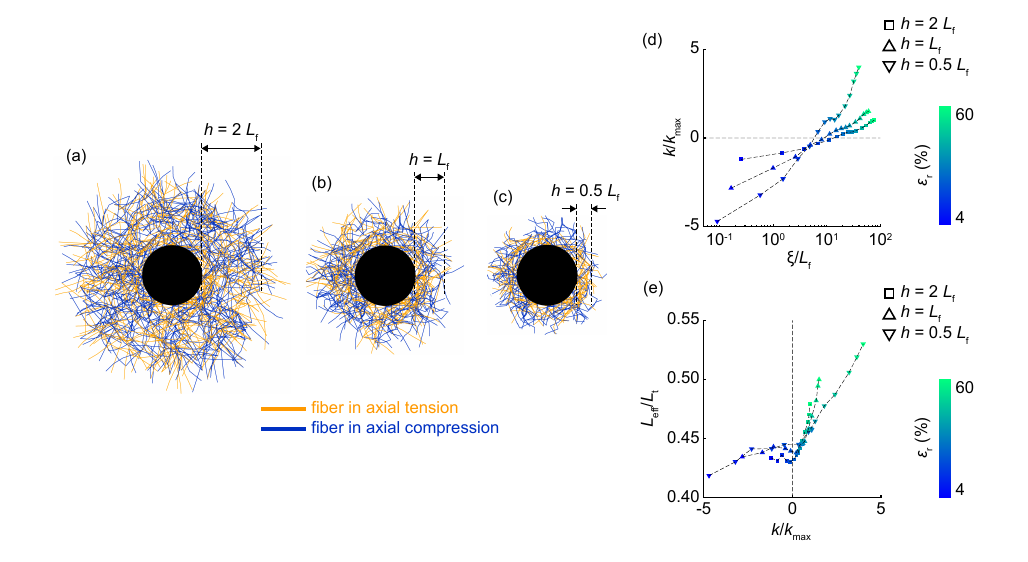}
\end{center}
\vspace{-16pt plus 2pt minus 2pt}
\textbf{Figure S6.}
Examining size effect on stiffness of remodeled matrix. 
(a--c) Representative annular regions with varying thicknesses, depicting three different matrix sizes, around a recovered inclusion (after $60\%$ pre-recovery inclusion contraction): $h=2L_\mathrm{f}$ (panel a), $h=L_\mathrm{f}$ (panel b), and $h=0.5L_\mathrm{f}$ (panel c). Tension and compression fibers are color-coded in orange and blue, respectively. Length dimensions 
in panels a–c are referenced in the Lagrangian coordinate system for the permanently deformed matrix regions around the recovered inclusion.
(d) Matrix stiffness depends on the total length of permanent defects ($\xi$) formed, showing softening ($k<0$) or stiffening ($k>0$) across different sizes as illustrated in panels a--c. Dashed lines connect individual data points to visually differentiate data from the three matrix size cases. $\xi$ was non-dimensionalized by the average matrix fiber length $L_\mathrm{f}$.
(e) Increase in the proportion of tension fibers ($L_\mathrm{eff}/L_\mathrm{t}$) as matrix stiffened ($k>0$) for all cases. Dashed lines connect data points to visually demarcate data from three scenarios.
Each data point in panels c and d corresponds to means from eight independent matrices.
Color bars adjacent to panels c and d indicate inclusion contraction levels ($\epsilon_\mathrm{r}$) for data points.

\section*{Supplemental Note 1: Rationale for Using a Contracting Inclusion to Apply Local Tension in Matrices}
In this study, we applied local tension within the matrix by using a contracting inclusion connected to several matrix fibers at their periphery, rather than pulling on individual fibers. This approach aligns with the goal of the manuscript to elucidate the geometry-driven mechanical memory formation in fibrous matrices, influenced by the collective assembly of fibers. This focus on matrix mechanics above the fiber length scales has been a foundational paradigm for understanding matrix behavior over the past two decades (e.g., Ref. \cite{picu2020mechanics}). It has also aided in deciphering mechanical cell-matrix interactions in biological processes such as wound healing and fibrosis, where contracting cells typically interact with multiple matrix fibers at length scales exceeding the average fiber length of the surrounding matrix. Prompted by these considerations, we strategically selected an inclusion diameter twice the average matrix fiber length to understand the mechanisms of mechanical memory formation driven by matrix geometry. While this manuscript delves deeply into the mechanisms underlying mechanical memory formation, our findings may have broader applicability. Expanding upon established trends of nonlinear matrix mechanics \cite{sarkar2024unexpected}, we cautiously speculate that the trends observed in this study could extend to all inclusion diameters exceeding the average fiber length.

\section*{Supplemental Note 2: Our Model Reproduces Foundational Insights}
The evolution of the areal density of defects produced by inclusion contraction ($\psi_\mathrm{1}$, Fig. 2c) suggests that at all levels of inclusion contraction, $\psi_\mathrm{1}$ decreased by over $85\%$ beyond $r=2L_\mathrm{f}$, highlighting that most defects occurred within $2L_\mathrm{f}$ from the inclusion. This zone, designated as the region of interest (Supplemental Fig. S1a), was chosen to reproduce the mechanical properties predicted in prior studies.

Within this region of interest (Supplemental Fig. S1b), two outcomes emerged as a result of defect formation, both consistent with prior findings \cite{grekas2021cells, proestaki2019modulus}. First, the interaction between buckled circumferential fibers and stretched radial fibers led to notable matrix densification ($\rho/\rho_\mathrm{ref}>1$, Supplemental Fig. S1c), with densification intensifying as the level of radial contraction increases. Second, a nonlinear mechanical response was observed in the matrix (Supplemental Fig. S1d), exhibiting softening ($k<0$) at lower radial strains and stiffening ($k>0$) at higher strains ($\epsilon_\mathrm{r}>50\%$). 

While increasing inclusion contraction led to expected densification and nonlinearity in the mechanical response of the matrix, it also revealed that at lower to intermediate contraction levels, the matrix softened despite an increase in density. This behavior aligns with recent findings \cite{sarkar2024unexpected,zakharov2024clots}, highlighting a peculiarity of fibrous materials where density alone does not determine stiffness. Instead, the evolving mechanical stiffness depends on the stress state of fibers. The matrix softens if buckled fibers dominate, even with densification at lower to intermediate levels of inclusion contraction. Consequently, density serves as a physical marker but cannot reliably indicate mechanical stiffness, emphasizing the pivotal role of fiber stress states in defining the mechanical response of the matrix.

\end{document}